\documentclass[12pt]{iopart}
\usepackage{epsfig}

\begin{document}

\title{Conventions spreading in open-ended systems} 

\author{E. Brigatti $^{\star,\dag}$ and I. Roditi$^{\star}$}  
  
\address{$^{\star}$Centro Brasileiro de Pesquisas F\'isicas and National Institute of Science and Technology for
Complex Systems,
Rua Dr. Xavier Sigaud 150,
22290-180, Rio de Janeiro, RJ, Brazil}
\address{$\dag$Instituto de F\'{\i}sica, Universidade Federal Fluminense, 
Campus da Praia Vermelha, 24210-340, Niter\'oi, RJ, Brazil}
\ead{ edgardo@cbpf.br}

\date{\today}
  
\begin{abstract}
We introduce a simple open-ended model which describes
the emergence of a shared vocabulary. 
The ordering transition towards consensus is generated only 
by an agreement mechanism. 
This interaction defines a finite and small 
number of states, despite each individual having the ability to 
invent an unlimited number of new words. 
The existence of a phase transition is studied, analyzing 
the convergence times, the cognitive efforts of the agents 
and the scaling behaviour in memory and time.  
\end{abstract}
    

\maketitle


\section{Introduction}

The general process where domains characterized by a specific 
homogeneous quantity emerge out of an initial disordered state
is a paradigmatic problem in Statistical Physics. 
A classical example is the 
study of magnetization, the Ising model 
being the simplest and more common representation 
of such phenomenon.

Interestingly, it is possible to redirect to this kind 
of situation problems 
coming from other areas of knowledge \cite{sociophysics}.
The spreading of opinions or conventions in social systems 
is a good example.
The analogy, in this case, can be built up through the identification 
of individuals with agents characterized 
by some quantity, and representing the interaction between 
them as a results of the implementation of a set of simple rules.
A straightforward realization of this scheme is possible using  
models inspired by the kinetic Ising model.
In this case, agents' opinions are characterized by the 
spin values and the interaction reduces to spin-flips, according  to 
some particular rule. 
For example, in two-state models, the opinion update (a spin-flip)
is implemented 
following the local majority \cite{galam,redner} or 
directly emulating another opinion \cite{voter,sznajd}.

Starting from these straightforward models, various levels of complexity have been 
reached with the introduction, for example, of noise \cite{borghesi} or 
with the implementation of a q-state discrete space (Potts-type models) \cite{slanina}.
A particularly interesting case was obtained coupling some 
Potts-type models \cite{axelrod}.
Finally, also quite different modelling strategy, which departs 
from the classical Ising prototype, appeared \cite{veit}.
\\


The specific situation where agents' opinions represent 
words and the social dynamics describes the evolution towards 
the emergence of a shared vocabulary, 
raised some novel questions that forced 
towards different modelling schemes .
In fact, these dynamics are characterized by 
some peculiar facts \cite{lingua}.
Linguistic units, because of their conventional nature \cite{witt}, 
are characterized by a strong arbitrariness, either in their 
changes or in their established forms.
This open-ended nature is constrained by the correlation 
imposed by communication:
the freedom of individual's inventiveness is bounded by 
the interaction with the community.
The effect of these social interactions is to reduce the arbitrariness 
of the used symbols and to convey the 
community towards a consensus.
Ultimately,  it is necessary to built up a connexion between these 
two founding opposing factors: the arbitrary convention, which allows 
totally free choice, and the social interactions 
during the passage of time, which fix that choice \cite{sauss}.\\

These last considerations suggest us to abandon the Ising-like 
canon. 
First of all, arbitrariness and variation in linguistic units 
force towards a model scheme
implemented in an open-ended word system.
These means that agents must be characterized by
a memory capable of accumulating an unlimited number of 
possible states, each one characterized by an infinite 
number of possible choices. 
Moreover, individuals must be able to constantly create new different words.
Second, the interaction between individuals should be 
characterized by rules which capture two essential features
of this process: the presence 
of memory and learning/feedback effects.
Unfortunately,  the herding behaviour described by
Ising-like interaction can not account for these effects.

This purpose has been recently achieved 
by the so called Naming Game \cite{baronca}, where the implemented 
microscopic mechanisms, 
the negotiation dynamics, are based on 
these features. 
In this dynamical system, originally
inspired by an artificial experiment based on robots \cite{steels},
the negotiation dynamics are able to order the system towards consensus 
without the intervention of any central control
and revealed itself richer and more realistic than the 
classical implementation based on Ising-like formalization.
\\ 

 
In the present work we will describe, throughout 
a modification of the Naming Game of reference \cite{baronca}, 
an  open system where each agent can actually store 
an unlimited number of different names. This  
fact is possible thanks to a new dynamics which allows
the introduction of different words at every M.C. step \cite{edo},
contrasting with a dynamics that
 allows the introduction of  different words just in the first M.C. steps \cite{baronca}.

With this setting, we investigate a minimal interaction mechanism 
capable of generating the convergence towards consensus.
In the original implementation two cognitive mechanisms  work:
the agreement and the learning (or overlapping) mechanism.
The first one erases the information collected in agent's 
memory, if it does not take part in the last agent's successful communication. 
The second one, in the case of a failure, allows the 
memorization of a new word, directly learned from the speaker. 
This last mechanism causes a fast increase in the overlap 
between the words used in the community  \cite{baronca,edo}.

In our work we looked for a lower bound for the cognitive 
mechanisms, defining a dynamics governed just by the 
agreement mechanism, with no internal overlapping mechanism.
In this situation all the new words collected in individuals' memory 
are self-invented, picked up by an external reservoir of words.
In  \cite{baronca}, after a transient, 
when everybody attains at least one word, 
the game is characterized by a fixed number of different 
words which are interchanged throughout the overlapping mechanism and are reduced by the agreement mechanism. In our model, 
new words are continuously invented
and the convergence does not simply emerge like an effect of 
the dissemination  of the most common word caused by the overlapping mechanism. 
In this way, we test the new hypothesis that it is possible to reach consensus even for memories which can store an infinite number of different words and with a simple interaction which can only compare and not exchange symbols.\\ 

In this implementation the nature of the words introduced in the
system is totally independent from the social mechanisms
of interaction.
Anyway, the nature of the linguistic convention is not totally 
arbitrary, and relates to society' s communicational use of the 
material supplied by the current language. 
Continuity with the past constantly restricts freedom of choice \cite{sauss}. 
This fact is a source of bounds in the open-ended nature of the
invented signs.
These considerations, summed to other practical reasons, suggested to characterize 
the reservoir of words by a distribution which models 
the presence of some constraints, differentiating between more and less probable words.\\

To sum up, in this work we will try to understand some basic features
of ordering processes governed by an agreement dynamics
with an unlimited number of memory states.
The importance of these dynamics is evident
in the description of conventions spread,
as, for example, in the convergence on the use of a 
particular word or category in real communities or 
among embodied software agents or tagging systems 
on the web.
The fundamental process acting in these dynamics is 
a memory-based negotiation strategy, made up by a sequence of 
trials which shape and reshape the system memories.
This kind of interaction is at the heart
of numerous cognitive processes, 
either in biological \cite{categories}, social \cite{peer} and technological 
\cite{steelsB} context, and it is more widespread 
than the simple imitation mechanism which can be
described by Ising-like dynamics.
For these reasons, the definition of
the simplest mechanism sufficient for its performance
is fundamental.
In this study, we will show that the agreement mechanism 
can be sufficient for reaching consensus and no other mechanisms,
which can help in diffusing the most common symbols, are necessary.
This fact is important not only from a theoretical point of view,
but also for a possible transposition into practical applications, as,
for example, the designing of artificial communication systems,
which can be represented either by robots \cite{steelsB} or by peer-to-peer 
information systems on the web \cite{peer}.



\section{The model}

The Game is played by $P$ agents. An inventory, which can 
contain an arbitrary number of words, represents each agent.
Population starts with empty inventories.
At each time step, two agents are randomly selected;
the first one assumes the role of speaker, the second one 
of hearer.
 Then, the following microscopic rules, which are a modification 
of the ones used in the original Naming Game \cite{baronca},  control their actions:

1) The speaker retrieves a word 
from its inventory or, if its inventory is empty, invents a new word.

2) The speaker transmits the selected word to the hearer.

3a) If the hearer's inventory contains such a word, 
the communication is a success. 
The two agents update their inventories so as to keep 
only the word involved in the interaction. 
 
3b) Otherwise the communication is a failure. 
The speaker invents a new word, 
different from all the other ones that keeps in its memory.\\

The players invent new words from a distribution that 
decays following a simple Gaussian law: $\exp(-x^{2}/2\sigma^{2})$. 
We choose this continuous distribution because its implementation
is easy and it allows fast simulation runs.
Anyway, words are represented only by positive integer numbers.
 To sum up, invention of new words corresponds to
pick them up from a reservoir, whose effective size is tuned 
by $\sigma$.  
The speaker chooses a new word which is
different from all the other ones that keeps in its memory 
but which can be present in the memory of different agents 
or could have been present, at earlier time steps, in the 
same agent's  memory.
By the way, this model implementation could generate homonymy 
if the agents 
try to achieve consensus on a set of names 
to be used for signifying a number of different meanings (objects).
\\

In accordance with our aims, the dynamics are governed 
only by the agreement mechanism and
the procedure for inventing new words is independent from 
the social mechanisms of interaction.
This implementation defines an open-ended 
system where an unlimited number of words
can be invented. Players invent new words if their inventory is empty
(that happens only in the early stages of the simulation),
or if their communication is a failure. In fact, we can think that,
in real life, individuals which are not able to communicate are naturally 
led to look for new words. 
  
\section{Results}

We will describe the time evolution of our system looking at some  
usual global quantities \cite{baronca,edo}: the total number of words ($N_{tot}$) 
present in the population, the number of different words ($N_{dif}$) 
and the success rate ($S$), which measures an average rate of success 
in communications.\\

\begin{figure}[p]
\begin{center}
\includegraphics[width=0.5\textwidth, angle=0]{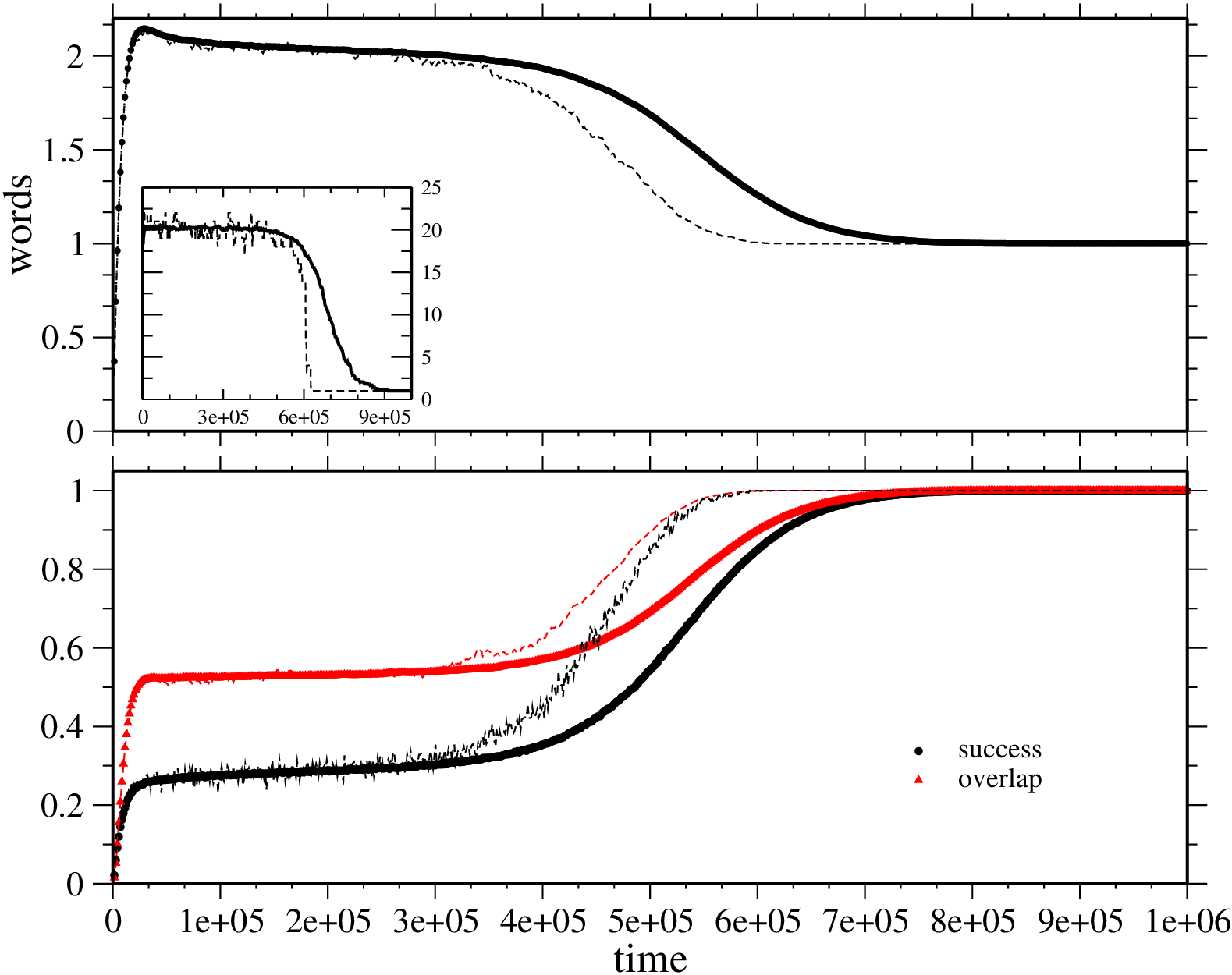}
\end{center}
\caption{\small Top: temporal evolution for the total number of words 
divided by the total population.
In the inset, total number of different words 
($N_{dif}(t)$). 
Bottom: the success rate ($S(t)$) and the overlap function $\mathcal{O}$. 
The dashed curves represent a single realization.
All other data are averaged over 100 simulations. ($P=7000$, $\sigma=5$). }
\label{fig_pheno}
\end{figure}

A fast initial transient exists. 
Agents start with an empty inventory and, 
in each interaction, each speaker invents at least 
a new word and each hearer can possibly learn one.
After this early stage,   the system self-organizes in a 
state where $N_{dif}$ and $S$  display a long plateau.
Similarly, the total 
number of distinct words displays a plateau, 
but slightly decreasing along the time evolution (see Figure~\ref{fig_pheno}).
Quite abruptly, perhaps because of a large fluctuation, 
the ordering transition take over.
The number of successful plays increases 
and $N_{tot}$ changes its concavity and begins a decay 
towards the consensus state, corresponding to 
one common word for all the players, reached at time $T_{c}$.
In contrast, the number of different words maintains a 
constant value until the convergence process has led the system 
nearer  to the absorbing state. 
In Figure~\ref{fig_pheno} we report the temporal evolution 
for $N_{tot}(t)$, 
$N_{dif}(t)$ and $S(t)$. 
Generally, the time evolution of the single 
runs displays a large variation and can be quite different from the 
average. \\


With the aim of characterizing with more details our system,
describing directly its state and not just the outcome of the game,
we looked at an overlap function \cite{baronca}. 
It corresponds to the total number of words common to all the 
possible agents' pairs:

\begin{eqnarray}
{\mathcal{O}}=\frac{2}{P(P-1)}\cdot \sum \limits_{i\ge j} a_{i}\bigcap a_{j}. \nonumber
\end{eqnarray}
The behavior of this quantity is similar to $S(t)$, 
with a well defined plateau
followed by a sudden transition towards one (see Figure~\ref{fig_pheno}).
The long plateau is characterized by a constant mean value 
which rapidly decreases increasing  the $\sigma$ value.
We will present a more comprehensive analysis of the behavior of this quantity at the end of this section.\\



We investigated how the macroscopic observables scale with 
the population size $P$. 

\begin{figure}[p]
\begin{center}
\includegraphics[width=0.5\textwidth, angle=0]{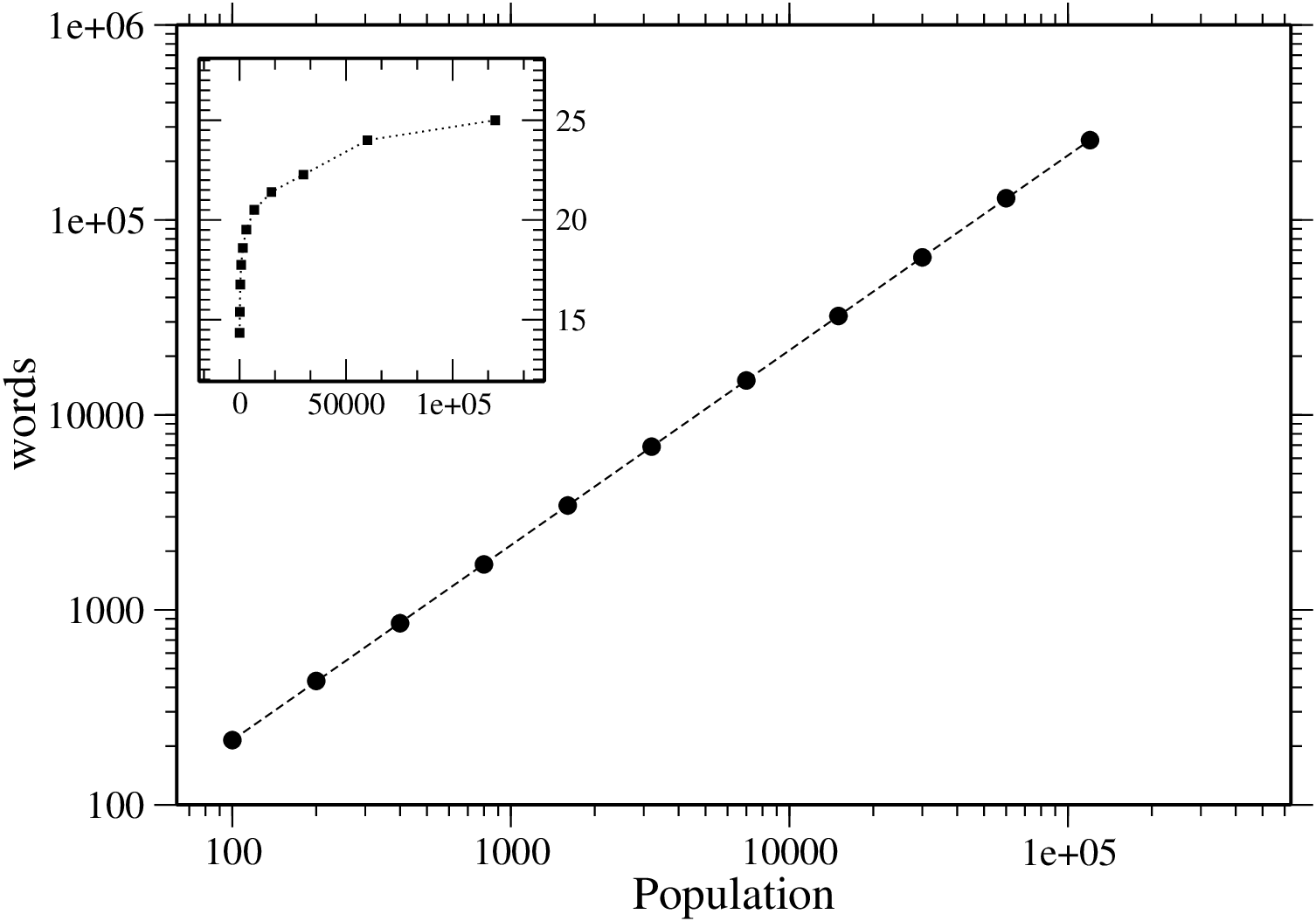}
\end{center}
\caption{\small Maximum number of total words (the dashed line has slope $1$) and, in the inset,
maximum number of different words for different population sizes.
In the original Naming Game \cite{baronca}, the maximum number of total words scales as $P^{1.5} $, and the
maximum number of different words scales as $P$. In our model
the first one scales as $P$ and the second one is almost not dependent on $P$, for sufficiently large $P$.
Data are averaged over 100 simulations with $\sigma=5$.}
\label{fig_pop}
\end{figure}

At first, we look at the scaling behavior of the system memory size.
By a numerical exploration, we can see that the maximum 
number of different words is almost not dependent on $P$, 
for sufficiently large $P$ (see Figure~\ref{fig_pop}).

The maximum number of total words 
linearly scales with the population number.
In other words, the number of total words of each 
player is not dependent on the dimension of the community.
 We can demonstrate it with a simple analytical 
consideration, already used in \cite{baronca}.
We represent  the mean total number of words for 
agent, at time step $t$, with $n(t)$, and  the mean total number 
of different words with $D(t)$.
If we assume that $n(t)$ scales as $\beta$, unknown, we can write:

\begin{eqnarray}
\label{eq_1}
n(t+1)-n(t)\propto \frac{1}{n(t)^{\beta}}\Big(1-\frac{n(t)^{\beta}}{D(t)}\Big)-
\frac{1}{n(t)^{\beta}}\frac{n(t)^{\beta}n(t)^{\beta}}{D(t)}
\end{eqnarray}
We are considering that the probability for the speaker to communicate 
a specific word is $\frac{1}{n(t)^{\beta}}$ and 
the probability for the hearer to own that words is $\frac{n(t)^{\beta}}{D(t)}$. 
It follows that the first term represents the gain term for a 
failed communication, and the second one the loss term 
We can use this equation for describing the $P$ dependence.
$D$, for large $P$, can be considered a constant. For this reason,
at the stationary state, where we should impose  $n(t+1)-n(t)=0$,
our equation reduces to $\frac{1}{n^{\beta}} \propto n^{\beta}$, which
forces $\beta=0$. This fact implies that 
the number of total words for each 
player is not dependent on $P$.
It follows that $N_{tot}\propto P $, as confirmed by the numerical 
results shown in Figure~\ref{fig_pop}.\\

\begin{figure}[p]
\begin{center}
\includegraphics[width=0.5\textwidth, angle=0]{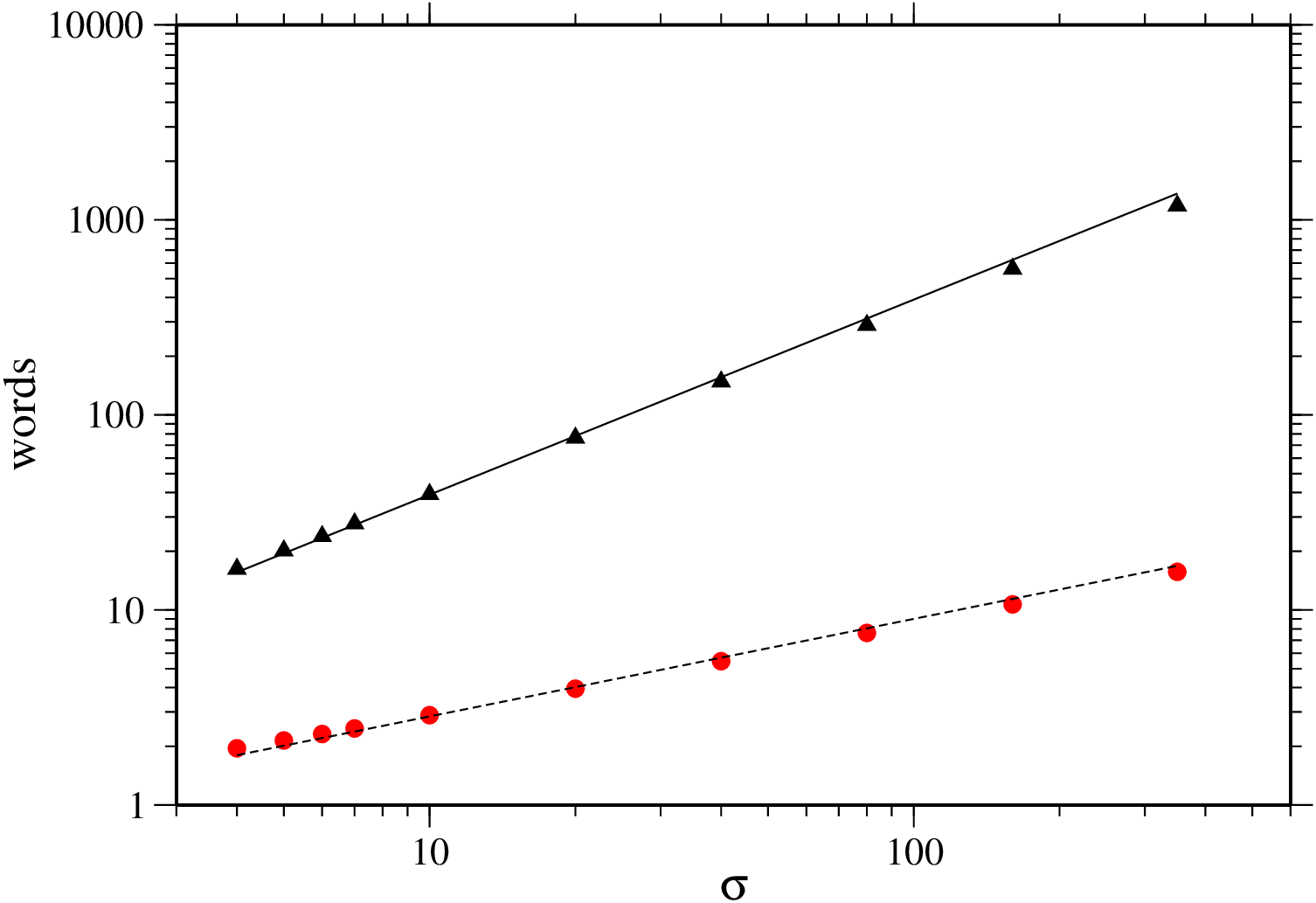}
\end{center}
\caption{\small Maximum number of total words (circles) and
maximum number of different words (triangles) for different $\sigma$ values.
The dashed line has slope $1/2$, the continuous one $1$.
Data are averaged over $100$ simulations with $P=5000$.}

\label{fig_sigma}
\end{figure}

We look at the behavior of our model varying the 
$\sigma$ value in the distribution of the invented words.
As can be seen in  Figure~\ref{fig_sigma}, $N_{dif.}$ 
linearly scales with $\sigma$.
This result is quite intuitive if we approximate 
our distribution of new words with a box of dimension $\sigma$.

We can use another time equation~\ref{eq_1} for examining
the dependence of $N_{tot}$ on  $\sigma$.
We will look for a power-law dependence on $\sigma$;
for this reason we consider that $n \propto {\sigma}^{\gamma}$
with $\gamma$ unknown.
Using the fact that, at equilibrium, $D \propto \sigma$,
we obtain:
\begin{eqnarray}
\frac{1}{\sigma^{\gamma}}\Big(1-\frac{\sigma^{\gamma}}{\sigma}\Big) \propto
\frac{1}{\sigma^{\gamma}}\frac{\sigma^{\gamma}\sigma^{\gamma}}{\sigma} \nonumber
\end{eqnarray}
This equation implies that, at the stationary state, 
$\gamma=1/2$. It follows that $N_{tot}\propto \sigma^{1/2}$, 
a result confirmed by the numerical data 
shown in Figure~\ref{fig_sigma}.\\

Finally, we explored the behavior of the convergence 
time $T_{c}$. As stated before, this is the time at 
which the system reaches 
the consensus state, corresponding to 
one shared word for all the players. When the system does not achieve convergence, the system persists in the plateau regime displayed in Figure\ref{fig_pheno}. 
There, each agent stores in its memory  some words 
which are continuously suppressed by successful 
communications and invented by failed communications.

\begin{figure}[p]
\begin{center}
\includegraphics[width=0.5\textwidth, angle=0]{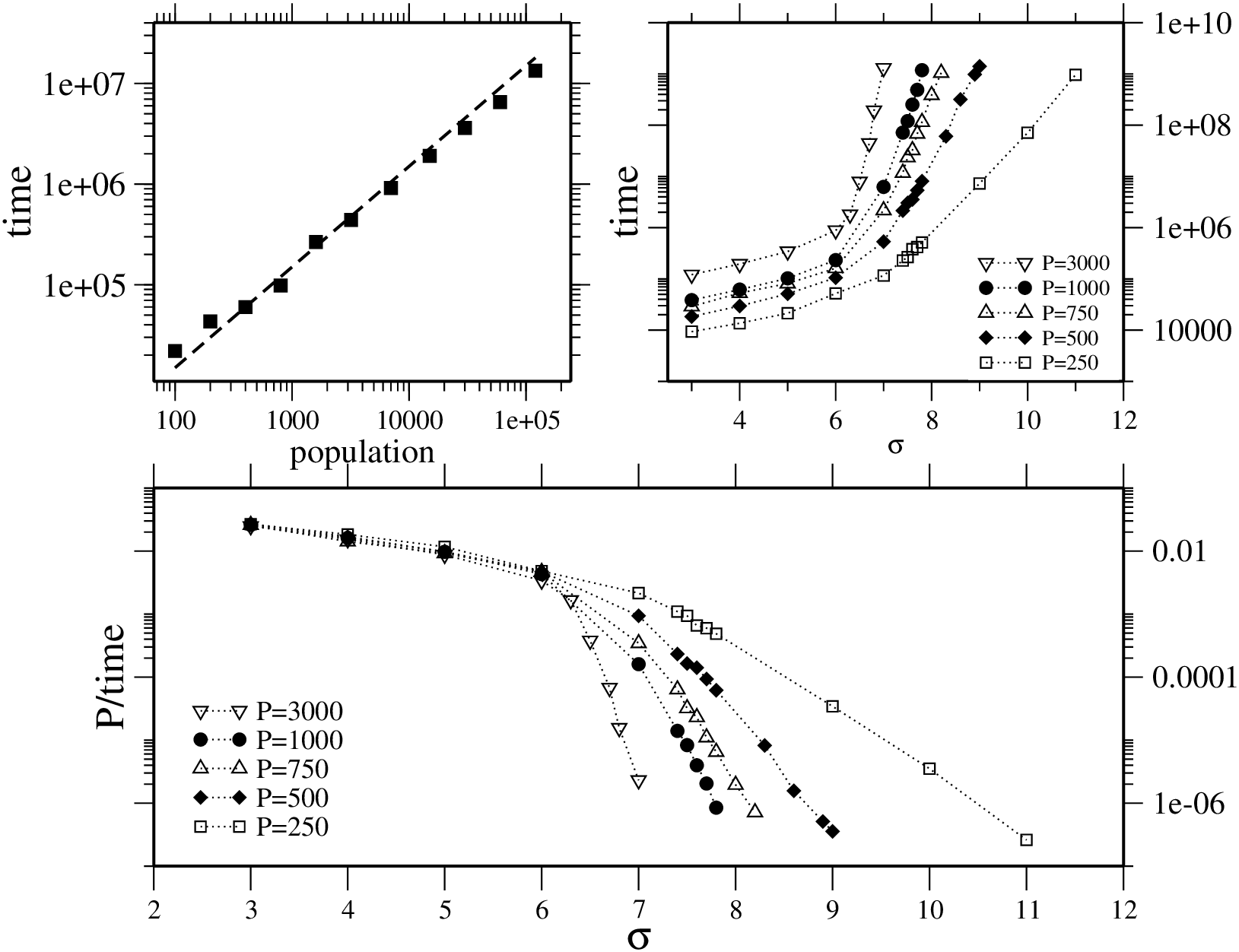}
\end{center}
\caption{\small Top: On the left, convergence time (convergence of the mean simulation) 
for different population sizes ( $\sigma=5$).
The dashed line has slope $1$. On the right, mean convergence time  for different  $\sigma$ values and different $P$. 
Bottom: Normalized inverse of the convergence time for different  $\sigma$ values and different $P$. }
\label{fig_tcritic}
\end{figure}

We stated the dependence of $T_{c}$ on $P$ averaging over 
different simulations.  
The $T_{c}$ value linearly grows with $P$ (see Figure~\ref{fig_tcritic}).
We have also observed that 
the distribution of the convergence times follows 
a log-normal like shape with large variances.

The $T_{c}$ dependence on $\sigma$ has a more intriguing behavior.
We report  $T_{c}$ as a function of the parameter $\sigma$,  
for different $P$ values (see Figure~\ref{fig_tcritic}). 
We display simulations up to values of $\sigma$ for which 
convergence emerges in the limits of typical computational 
times ($2\times10^{9}$ steps).
For low $\sigma$ values, $T_{c}$ increases exponentially.
For higher values, the divergence gets super-exponential
and the linear scaling, as a function of $P$, breaks down.
By increasing the system size, 
the growth of $T_{c}$ becomes steeper.
Anyway, it is difficult to state if exists a $\sigma_{crit}$ which corresponds
to a phase transition between a stationary state characterized by consensus and one 
characterized by disorder.
The plot of $(T_{c})^{-1}$ can suggest this fact (see Figure~\ref{fig_tcritic}), with
$6<\sigma_{crit}<7$ for large $P$.
Unfortunately, because of the open-ended nature of our model,  analytical results, 
which can confirm  this supposition, are still lacking.
An interesting system where, even though with a different
mechanism, an analogue transition has been described also analytically,
can be find in  \cite{baronca1}. There, a consensus phase transition, controlled by a 
parameter which represents the efficiency of the negotiation dynamics, 
is described. This parameter constrains the level of noise present in 
communications (no noise corresponds to the Naming Game). For a sufficient 
low level of noise, consensus is reached. Otherwise, several opinions, continuously exchanged between agents, coexist.
There is a similarity between increasing the noise level,
which decreases the efficiency of the erasing mechanism, and increasing 
the broadness of our reservoir, which corresponds to increase the variance 
of the words entering the system.

We conclude this analysis comparing our results 
with the ones of the original Naming Game \cite{baronca}. 
In the thermodynamic limit, the usual Naming Game stores an infinite 
number of different words ($P$-dependence), and, also after the temporal 
rescaling $t\to t/P$, the convergence time diverges ($P^{1.5}$-dependence).
In our model, in the thermodynamic limit, the number of different words 
seems to be finite and, at least for  $\sigma<\sigma_{crit}$, the system can converge 
towards consensus ($P$-dependence).
Now, for $\sigma\to\infty$, the number of words becomes infinite and 
consensus is not attained.
This fact suggests that, if the number of words effectively presents 
in the system is infinite, convergence can not be attained.\\

Finally, we study the dependence of the convergence towards 
consensus on the initial conditions of our simulations.
The overlap function ${\mathcal{O}}$ is the one which can 
better represent the system internal state.
For this reason, we characterize the initial conditions
with their ${\mathcal{O}}$ value.
We obtain an ensemble of different initial conditions
running a simulation with some fixed parameters and
picking up some particular configurations obtained at a 
specific time step.
Then, we run new simulations, with different parameters, 
starting from these collected configurations.

\begin{figure}[p]
\begin{center}
\includegraphics[width=0.5\textwidth, angle=0]{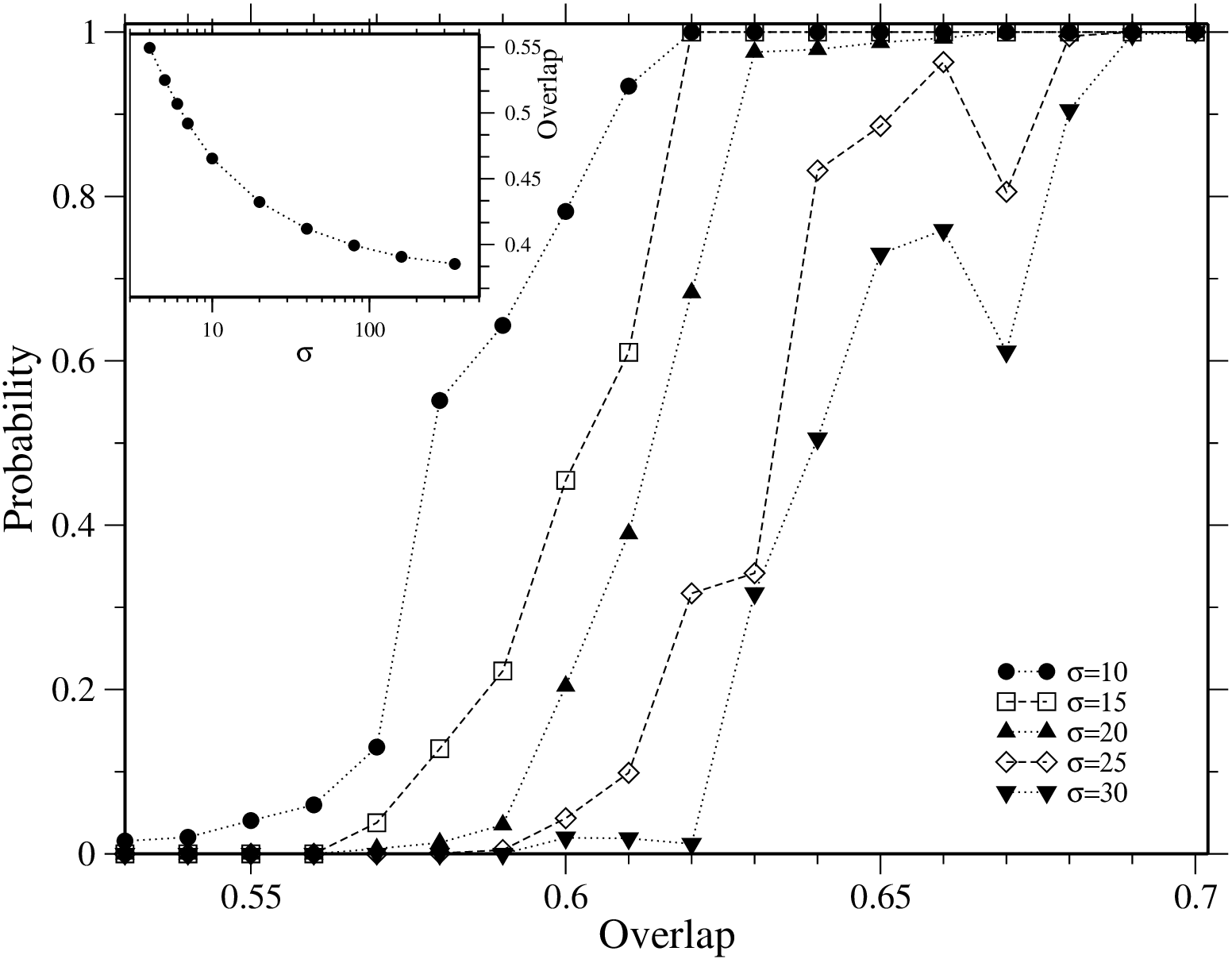}
\end{center}
\caption{\small  Probability of reaching consensus 
as a function of the ${\mathcal{O}}$ value of the 
initial conditions, for different $\sigma$ values ($P=1000$). 
In the inset, mean value of ${\mathcal{O}}$ in the plateau region, 
as a function of $\sigma$, for standard initial condition.}
\label{fig_prob}
\end{figure}

In Figure~\ref{fig_prob} we report the frequency of runs which 
reach consensus as a function of the ${\mathcal{O}}$ value of the 
initial conditions, for different $\sigma$ values. 
The probability of reaching the ordered state
increases with higher ${\mathcal{O}}$ values.
The same probability decreases by rising the 
$\sigma$ value of the run.
In fact, higher $\sigma$ values cause a slower ordering or even do 
not allow the transition to take over.
With this approach is easier to determine if 
the system reaches consensus or not, without having to 
wait for very long time.
In fact, we consider that the system does not evolve towards consensus if
the initial overlap value decreases until reaching the stationary 
state value corresponding to that $\sigma$.
We consider that the stationary ${\mathcal{O}}$ value 
(a function of $\sigma$)
is the one previously found at the end of very long simulations run 
with standard initial conditions (see inset of Figure~\ref{fig_prob}) .

A simple interpretation of these results is possible.
A critical overlap value exists: when the system
reaches that value the ordering transition takes place and a 
very fast convergence process toward the absorbing state occurs.
The convergence probability corresponds to the probability that, 
because of a fluctuation, such critical value is achieved 
starting from the ${\mathcal{O}}$ value of the plateau. 
For this reason, different initial conditions in the ${\mathcal{O}}$ value 
(${\mathcal{O}}\neq 0$) cause the convergence for $\sigma$ values which 
do not allow convergence for initial configuration having ${\mathcal{O}}=0$. 
Moreover, such critical ${\mathcal{O}}$ value grows with $\sigma$ and
this is the reason why the convergence is not always attained.


\section{Conclusions}

 We propose a very simple model,  a modification of the 
Naming Game of reference  \cite{baronca},  which 
explores some questions raised by the study of
semiotic dynamics.
 We implement an open-ended systems, where
only the agreement mechanism is responsible 
for the ordering process.
This interaction is sufficient for defining a finite and small 
number of states, despite each individual having the ability to 
invent an unlimited number of new words. 
We found that, for some values of the parameters, 
the ordering process led to a consensus state, 
characterized by the use of a single word.
With these results we confirmed the hypothesis that it is possible to reach consensus even for memories 
which can store an infinite number of different words and with a simple interaction which can only compare and not exchange symbols. This is a quite unexpected result. In fact, if exchanging symbols in a closed systems can quite intuitively lead to convergence, it would be 
hard to foresee the same behavior by simply erasing symbols in an open system.
These facts are explored throughout an extended 
numerical study which
analyzes the scaling behavior for memory and time, with respect to the population
size.
Moreover, we study the relation between convergence times and shape of
the invented words distribution and the dependence on initial conditions.

Our analysis assesses the robustness of the 
ordering transition
for models with a very simple interaction mechanism
and gives some new general insights for systems characterized by 
an unlimited number of memory states.
It describes some characteristics of the birth of neologisms
in real communities, modelled through the introduction of a reservoir of words which reflects the fact that the nature of the 
linguistic convention is not totally arbitrary. 
These results, interesting from a theoretical point of view, 
perhaps could be transposed into practical applications for 
designing artificial communication systems for robots \cite{steelsB}
or for others semiotic dynamics \cite{peer}.


\section*{Acknowledgments} 
We thank the Brazilian agency CNPq and FAPERJ 
for financial support.

\end{document}